\documentclass[reprint,
 onecolumn,
 amsmath,amssymb,
 pra,
floatfix,
]{revtex4-2}
\usepackage{graphicx}
\usepackage{subcaption}
\usepackage{float}
\usepackage{dcolumn}
\usepackage{bm}
\usepackage[colorlinks,allcolors=blue]{hyperref}      
\usepackage[nameinlink,capitalize,noabbrev]{cleveref}

\begin{document}

\title{Dynamic Shock Recovery in IO Networks with Priority Constraints}

\author{Jichu Han}
\email[Author: ]
{hanjichu01@gmail.com}
\affiliation{University of California, Berkeley}
\author{Lina Wang}
\affiliation{University of California, Berkeley}
\author{Richard Bookstaber}
\affiliation{MSCI}
\author{Dhruv Sharma}
\email[Corresponding Author: ]{dhruv.sharma@msci.com}
\thanks{\\ \textbf{Authors’ Disclaimer}: The views, thoughts, and opinions expressed in this article are those of the authors alone and do not necessarily reflect the official policies, positions, or opinions of any current or former employers, affiliated institutions, organizations, or any other individuals or entities. This article is intended for informational purposes only and should not be construed as professional or legal advice. The authors bear sole responsibility for any errors or omissions.The views expressed in this article are those of the authors and do not necessarily reflect the positions of their affiliated institutions or employers.}
\affiliation{MSCI}

\date{\today}

\begin{abstract}
    Physical risks, such as droughts, floods, rising temperatures, earthquakes, infrastructure failures, and geopolitical conflicts, can ripple through global supply chains, raising costs, and constraining production across industries. Assessing these risks requires understanding not only their immediate effects, but also their cascading impacts. For example, a localized drought can disrupt the supply of critical raw materials such as cobalt or copper, affecting battery and electric vehicle production. Similarly, regional conflicts can impede cross-border trade, leading to broader economic consequences. Building on an existing model of simultaneous supply and demand shocks, we introduce a new propagation algorithm, Priority with Constraint, which modifies standard priority-based rationing by incorporating a minimum supply guarantee for all customers, regardless of their size or priority ranking. We also identify a buffer effect inherent in the Industry Proportional algorithm, which reflects real-world economic resilience. Finally, we extend the static shock propagation model to incorporate dynamic processes. We introduce mechanisms for gradual shock propagation, reflecting demand stickiness and the potential buffering role of inventories, and gradual recovery, modeling the simultaneous recovery of supply capacity and the inherent tendency for demand to return to pre-shock levels. Simulations demonstrate how the interplay between demand adjustment speed and supply recovery speed significantly influences the severity and duration of the economic impact after a shock.
\end{abstract}

\maketitle

\section{Input-Output Models}

During World War II, the Allied Forces sought to cripple the German war machine by identifying a single point of failure. For this task, they turned to an input-output model pioneered by a Russian-American economist, Wassily Leontief~\cite{leontief1936quantitative, leontief1986input}. By mapping Germany's industrial interdependencies, the analysis pinpointed ball bearings as indispensable for a wide range of military equipment, from tanks to aircraft. This insight prompted a series of high-risk bombing raids on the ball-bearing factories in Schweinfurt at a heavy cost; over 60 bombers were destroyed. Unfortunately, the prospects of an end to the war through these heroic missions were short-lived because the Germans adapted, rebuilding with decentralized production in many smaller facilities.

In more peaceful applications, input-output analysis reveals the interdependencies between industries: it shows how outputs from one sector serve as inputs for another, uncovering the structure of production. This tool remains invaluable for assessing supply-demand shocks, economic growth, and policy effects. It offers an understanding of how changes in one sector ripple through an entire economy. For a more recent discussion of the use of input-output analysis to understand post-Covid inflation dynamics, see~\cite{weber2024inflation}. 

To begin with, input-output analysis provides a framework for understanding the interdependency within an economy. It does so by mapping how the output of one industry serves as an input for another, revealing how industries rely on each other in a web of production. The goal of input-output analysis is to quantify how changes in demand or supply in one sector propagate through the economy, affecting other sectors directly and indirectly.

At the core of this analysis is the input-output matrix, often referred to as the $A$ matrix. Each element of this matrix, $a_{ij}$, represents the amount of output from industry $i$ that is required as an input for industry $j$ to produce one unit of output. In other words, the $A$ matrix captures the input requirements across industries, showing how the production in one sector depends on inputs from others.

The model typically utilizes the Leontief inverse, represented as $(I - A)^{-1}$, where $I$ is the identity matrix and $A$ is the input-output matrix. The inverse is used to determine the total effect of a change in demand, not just the direct effect but also the indirect (secondary, tertiary, and so on) effects across the economy. By applying the inverse, we capture how a shock in one sector results in cascading effects across the interconnected industries.

The power series expansion of the Leontief inverse illustrates how these effects accumulate. The inverse $(I - A)^{-1}$ can be expressed as an infinite series:

\[
(I - A)^{-1} = I + A + A^2 + A^3 + \dots
\]

In this expression, the identity matrix $I$ represents the direct impact, while higher powers of $A$ (such as $A^2$, $A^3$, etc.) correspond to the secondary, tertiary, and further-round effects. The term $A^2$ shows the impact of indirect, second-order effects, $A^3$ reflects third-order effects, and this pattern continues, capturing the complete ripple of impacts throughout the economy.

\section{An Input-Output Model with Both Supply and Demand Shocks}

The Leontief model is a foundational tool for analyzing economic interdependency. However, it has several important limitations. The Leontief input-output model is primarily demand-driven and struggles to account for supply-side disruptions. It assumes fixed production coefficients and cannot dynamically adjust to input shortages, logistical bottlenecks, or simultaneous supply shocks. This limitation becomes critical in today's interconnected global economy, where firms often rely on multiple inputs that may face distinct and concurrent constraints. As a result, the model fails to capture the compounded effects of supply-side shocks across complex supply chains.

In the paper~\cite{pichler2022simultaneous} authors Anton Pichler and J. Doyne Farmer address these limitations in traditional input-output (IO) models. These limitations became particularly relevant during the Covid-19 pandemic, where both supply constraints (due to lockdowns and sickness) and demand reductions (due to lower consumption) occurred concurrently. The paper introduces a dynamic approach to modeling these simultaneous shocks. The authors extend the IO framework to account for both types of constraints. Specifically, their approach highlights how different prioritization strategies (e.g., proportional, industry proportional, and priority-based rationing) influence the distribution of constrained resources and the subsequent economic impact across interconnected industries.


We detail the different rationing algorithms employed in \cref{app:ration_algorithm}. All the rationing algorithms share the same core idea --- calculating a constraint based on supply limitations. They differ in what type of demand is prioritized and how the resources are distributed. 
\begin{itemize}
    \item \textbf{Proportional}: Treats all customers equally, distributing resources based on the total demand without prioritization.
    \item \textbf{Industry proportional}: Prioritizes intermediate demand, ensuring that other industries receive necessary inputs before final consumers are served.
    \item \textbf{Priority}: Prioritizes based on customer size, serving the largest customers first and potentially leaving smaller ones with little to no supply.
\end{itemize}

These differences reflect the varying strategies for handling supply shortages, depending on whether fairness, industry production continuity, or economic power is prioritized.

The proportional algorithm directly takes into account the final consumer's demand. If the supplier cannot supply the final consumer, they are going to cut the supply. But for the industry proportional algorithm, the other industries are considered as demand constraints rather than final consumers. Taking a step further, the Priority algorithm shows how the industries supply the other industries. They tend to serve the largest customer first and then turn to smaller ones. Finally, priority with constraint algorithm limits the minimum supply of industries, enabling the smaller industries to still be served in the presence of the priority algorithm.

Thus, the combination of supply and demand shocks forces a more sophisticated approach to resource allocation. Unlike single-sided constraints, where market forces can adjust naturally, the interaction between constrained production and demand creates a situation where the methods of rationing—whether proportional, priority, or priority with constraint—directly shape the broader economic outcomes. Understanding these mechanisms is key to navigating crises where both supply and demand are disrupted simultaneously. 

\begin{figure}[h] 
    \centering
    \includegraphics[width=\textwidth]{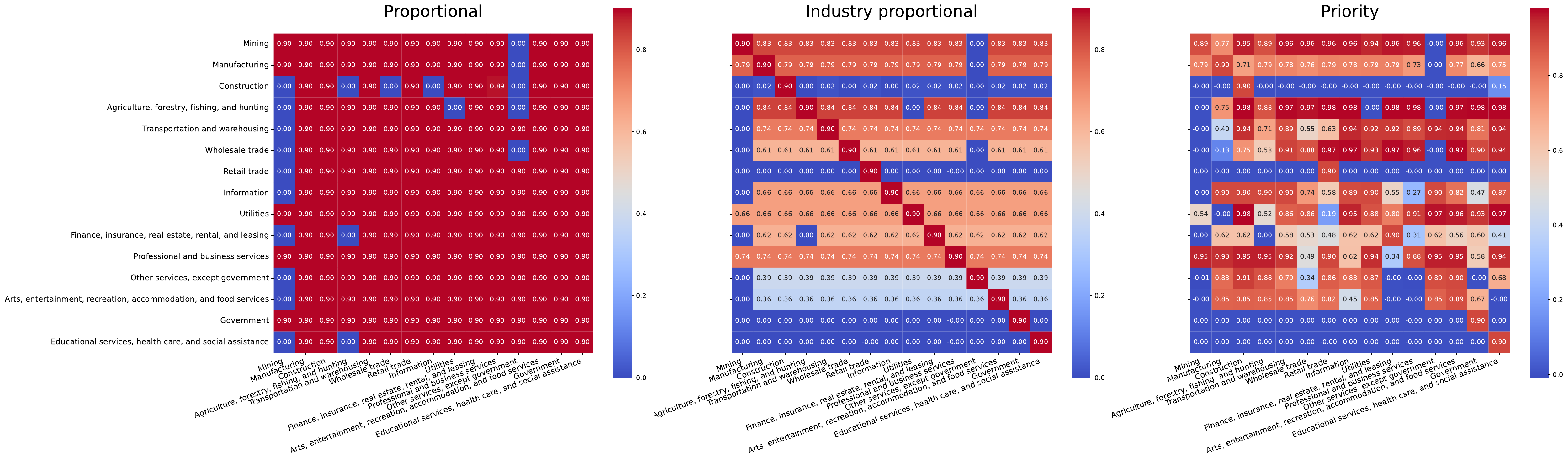} 
    \caption{\textit{Proportional, Industry proportional and priority algorithm impact matrices. Each cell shows the percentage reduction in final demand for industry j (column) resulting from a 90\% supply shock to industry i (row). Industries are sorted from upstream (top/left) to downstream (bottom/right). }}
    \label{fig:heat_map_first3}
\end{figure}

We show the impact of the propagation of a shock in each industry and compute the impact of that shock to all other industries through the impact matrix. The impact matrix quantifies the extent to which a supply shock in one industry affects all other industries. Each row represents the source industry experiencing the shock, while each column shows the affected industry. Specifically, the matrix entries indicate the percentage reduction in final demand for each column industry resulting from a given percentage supply shock in the row industry. In the heatmap shown in \cref{fig:heat_map_first3}, we illustrate the effect of a 90\% supply shock across industries. For example, the row corresponding to the Mining industry shows how a 90\% disruption in mining supply translates into percentage reductions in final demand across all other industries. We are using a limited subset of the complete IO Matrix from the Bureau of Economic Affairs(BEA). 

To better visualize the impact matrix, we sort industries from upstream to downstream. We define upstream industries as those with higher total final output, as they typically supply a larger share of inputs to other sectors. Conversely, downstream industries have lower final output and are more dependent on inputs from others. This ranking allows for a more intuitive interpretation of the propagation of supply shocks through the production network. This notion of upstreamness/downstreamness is similar to the definition provided in \cite{bartolucci2025upstreamness}.

In \cref{fig:heat_map_first3}, we present the results for the first three propagation algorithms introduced in the original paper. In the proportional algorithm, the initial shock propagates through the network without attenuation. This occurs because final consumer demand is incorporated into the bottleneck calculation of supply disruptions, resulting in a uniform bottleneck across all industries. The industry-proportional algorithm reveals heterogeneous impacts across industries. Shock on downstream industries experience less disruption on economy than upstream ones, as the demand for their outputs are less which makes it easier to meet. Finally, the priority algorithm exhibits even more idiosyncratic effects: larger buyers are better protected, while lower-priority customers face greater losses. These customers are vulnerable nodes in the network subjected to loss.

\vspace{-0.5cm}

\section{A new rationing algorithm - Priority with Constraint}

Building on this prior work, our paper introduces a novel rationing algorithm—Priority with Constraint. It ensures minimal supply to all industries regardless of priority, enhancing system robustness by preserving supply chains. In addition, it provides a buffer for lower-priority consumers, reflecting real-world conditions. For example, if a lower priority entity, such as the government, receives no mining supply, the impact is minimal due to its lower dependence on that sector. Minimal supply thus acts as a safeguard in downstream propagation.

\begin{figure}[H] 
    \centering
    \includegraphics[scale=0.4]{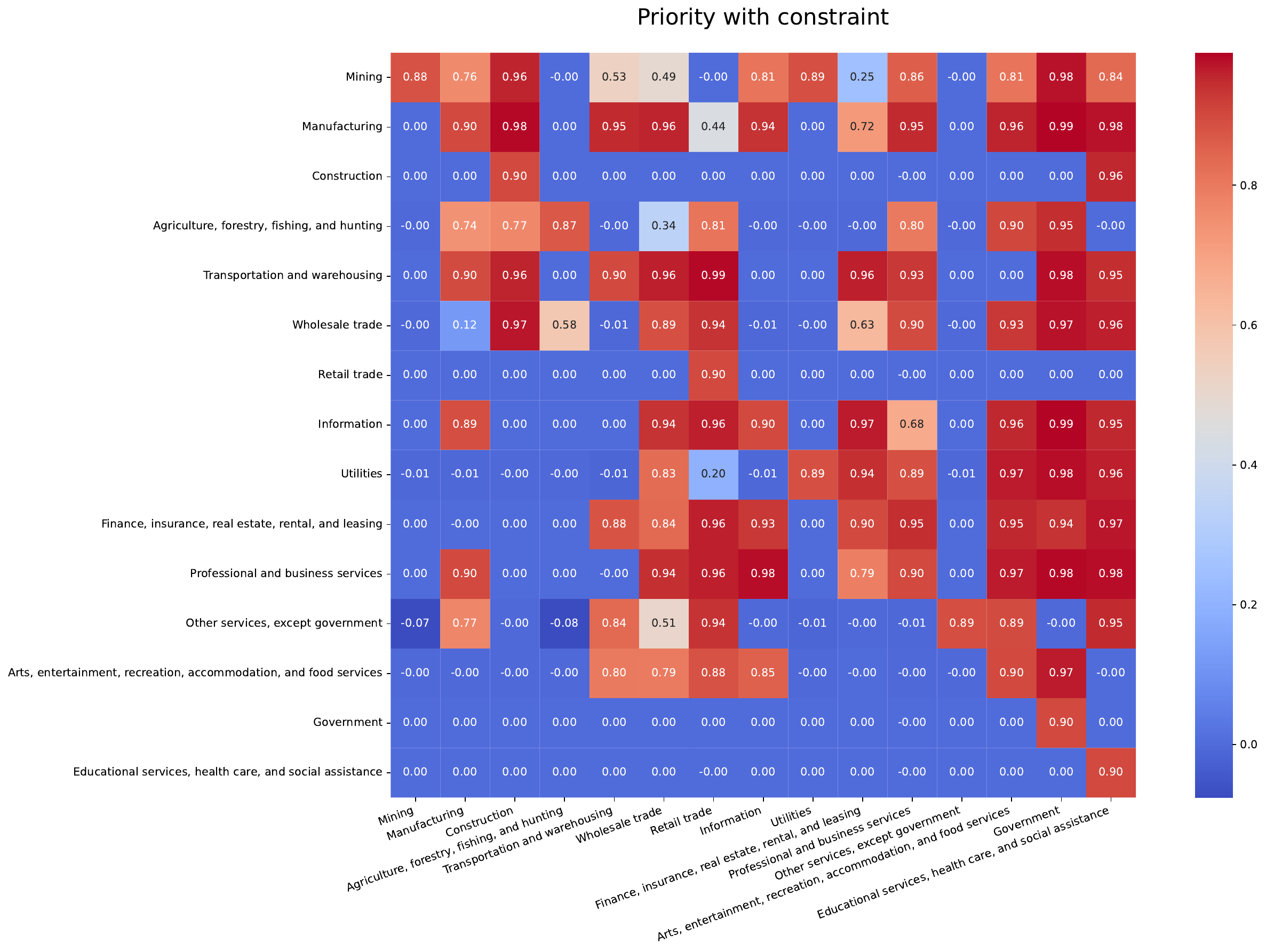} 
    \caption{\textit{Priority with Constraint impact matrix resulting from a 90\% supply shock. Note the attenuation compared to other algorithms and the concentration of impact in the upper triangle, indicating upstream industry influence.}}
    \label{fig:Priority with constraint}
\end{figure}

In \cref{fig:Priority with constraint}, we show the result for priority with constraint. The result showed further attenuation of shock, since there's a minimum supply of each industry. Meanwhile, it reveals that most of the impact concentrates in the upper triangular region, indicating that upstream industries exert greater influence on the economy, while downstream industries are less influential—consistent with economic intuition.

\section{Buffer Effect}

Apart from the Proportional algorithm, which imposes a uniform bottleneck across all industries, other algorithms incorporate mechanisms to absorb shocks, thereby creating a buffer within the economy. The mechanism is given by industry Proportional, where a supplier's bottleneck is determined by its capacity to supply intermediaries. Final consumer demand is treated as flexible, and the supplier is unlikely to restructure its output in response to a decline in final demand. As a result, when the supply shock is small, the supplier may still meet intermediary demand, despite reduced final consumption. In such cases, no binding bottleneck emerges, and the shock is absorbed by the final consumers without propagating through the production network.

The shock affects the economy only if it is sufficiently large. For instance, a severe disruption in the mining industry may reduce output enough to impact downstream sectors such as transportation and finance, potentially leading to broader economic consequences. However, such cases are rare. Most shocks are minor and can be internally absorbed—for example, through supply reallocation or inventory use—allowing the mining industry to minimize disruptions. In these cases, while the mining sector is affected, downstream industries may remain largely unaffected.

\begin{figure}[h] 
    \centering
    \includegraphics[scale=0.2]{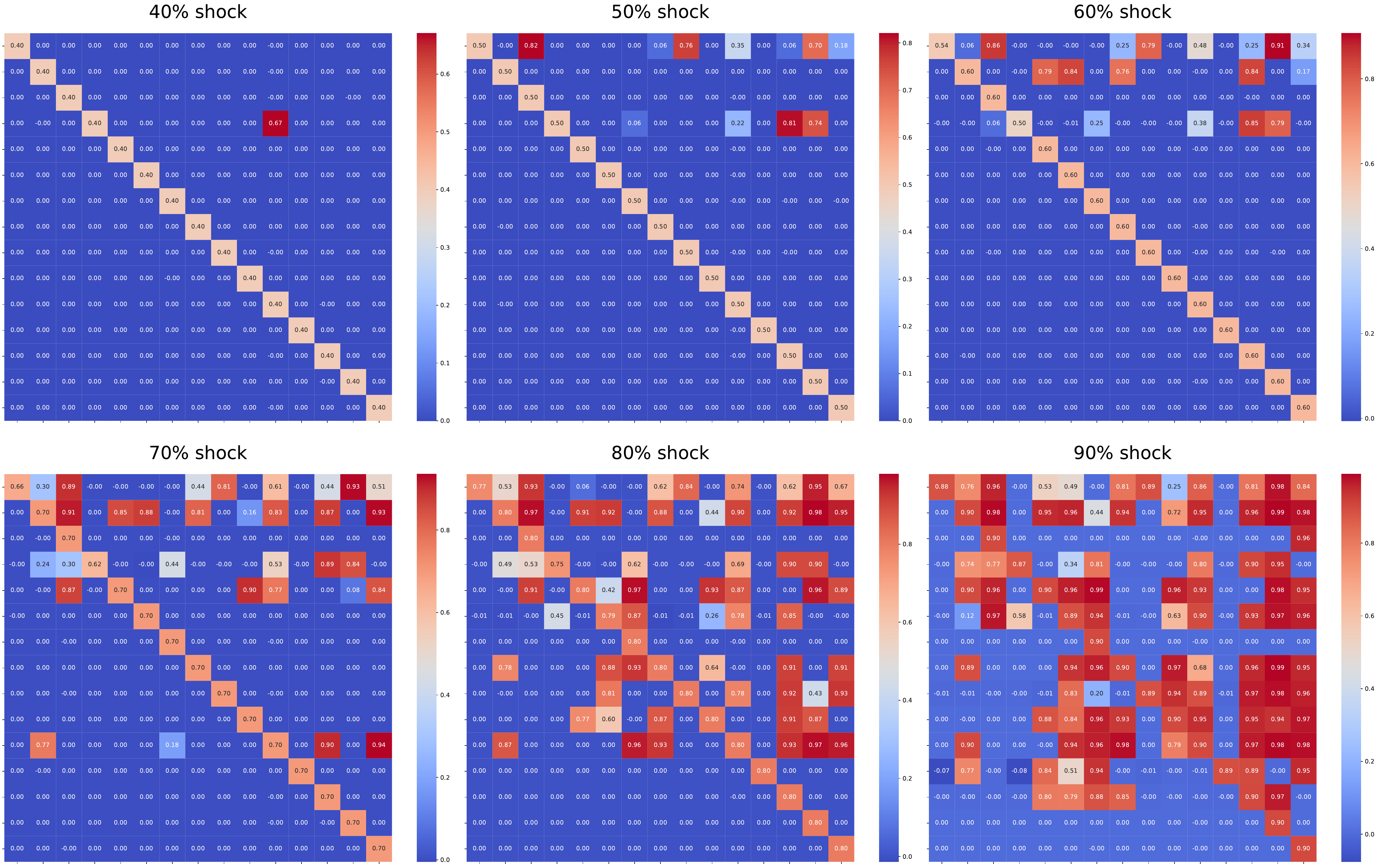} 
    \caption{
    \textit{Demonstration of the buffer effect using the Priority with Constraint algorithm. Each heatmap shows the cross-industry impact resulting from an initial supply shock of varying magnitudes (10\% to 90\%, indicated on the subtitles) applied individually to each source industry (rows). Widespread propagation across industries (significant off-diagonal impacts) only becomes apparent when the initial shock exceeds approximately 40\%.}}
    \label{fig:buffer_effect}
\end{figure}

We summarize this effect in \cref{fig:buffer_effect}. For a sequence of input shock values, we observe how the other industries are impacted. While local impacts are seen for a 40\% shock, we only start to observe the impact across a wide range of industries at higher levels of the shock.

\textbf{Note:} The magnitude of the shock in \cref{fig:Priority with constraint} and \cref{fig:buffer_effect} is proportional to the total output of each industry. However, the current model lacks appropriate adimensionalization, as the Input-Output (IO) matrix is expressed in units of dollars. Rescaling the IO matrix to a dimensionless form would allow for a more uniform analysis by ensuring that the relative impact of shocks is comparable across industries, regardless of their total output.

\section{Propagation and Recovery: Path-based modeling}

While the approach presented in \cite{pichler2022simultaneous} proposes a crucial extension to the traditional Leontief framework, they are silent on the recovery mechanism following a supply-demand shock. In what follows, we extend their model to introduce gradual propagation and recovery. We detail the propagation and recovery dynamics in Appendix~\ref{app:io_algorithm}. Similar recovery dynamics, following a Covid-19 like shock, can be seen in \cite{sharma2020v}. 
\begin{enumerate}
    \item 
\textbf{Gradual Propagation} \\
Subject to a supply disruption, we might expect demand to gradually drop since people cannot adjust immediately to the shock. There may be inventory in the economy that can provide some buffer for consumers. Therefore, we apply an exponential weighted moving average to the final consumer demand controlled by demand adjustment speed.

\item 
\textbf{Gradual Recovery: Recovery speed} \\
To enable the economy to recover, both supply and demand should recover simultaneously. Therefore, we assume that demand always has a motivation to recover to its original level, though it will not be fully satisfied due to existing bottlenecks. Meanwhile, the supply power will gradually recover, allowing demand to be satisfied and enabling the economy to return to its original level. To model this, we set expected demand at the original level as the target of recovery, and the satisfied demand is always lower than expected demand due to supply shock but has a tendency to recover to the target at recovery speed once supply power recovers.
\end{enumerate}

\begin{figure}[h]
    \centering
    \begin{subfigure}[b]{\textwidth}
        \centering
        \includegraphics[width=0.8\textwidth]{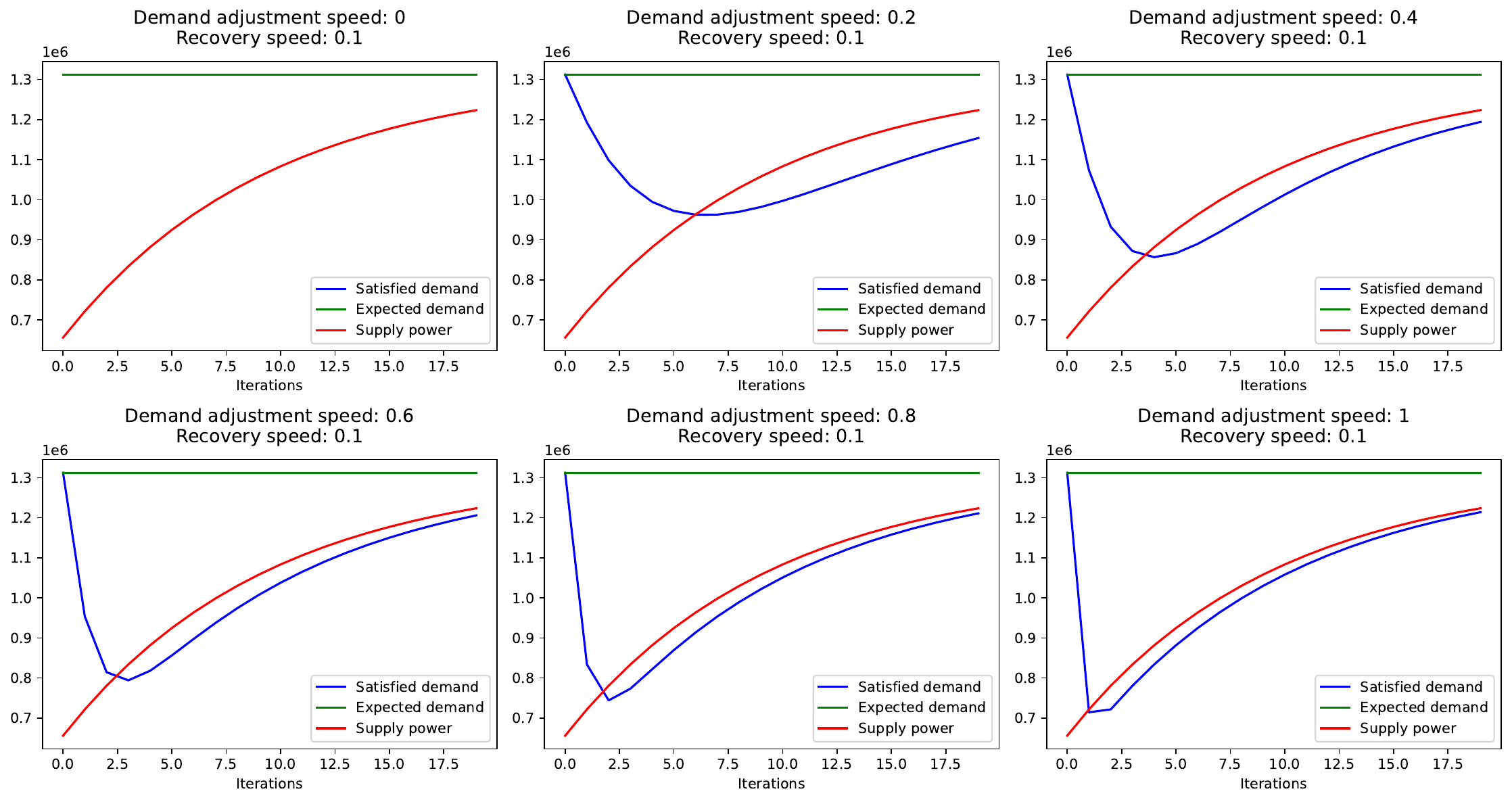}
        \caption{\textit{The impact of demand adjustment speed with a fixed supply recovery speed of 0.1}}\label{fig:recovery_1_sub}
    \end{subfigure}
    \vspace{1em}
    \begin{subfigure}[b]{\textwidth}
        \centering
        \includegraphics[width=0.8\textwidth]{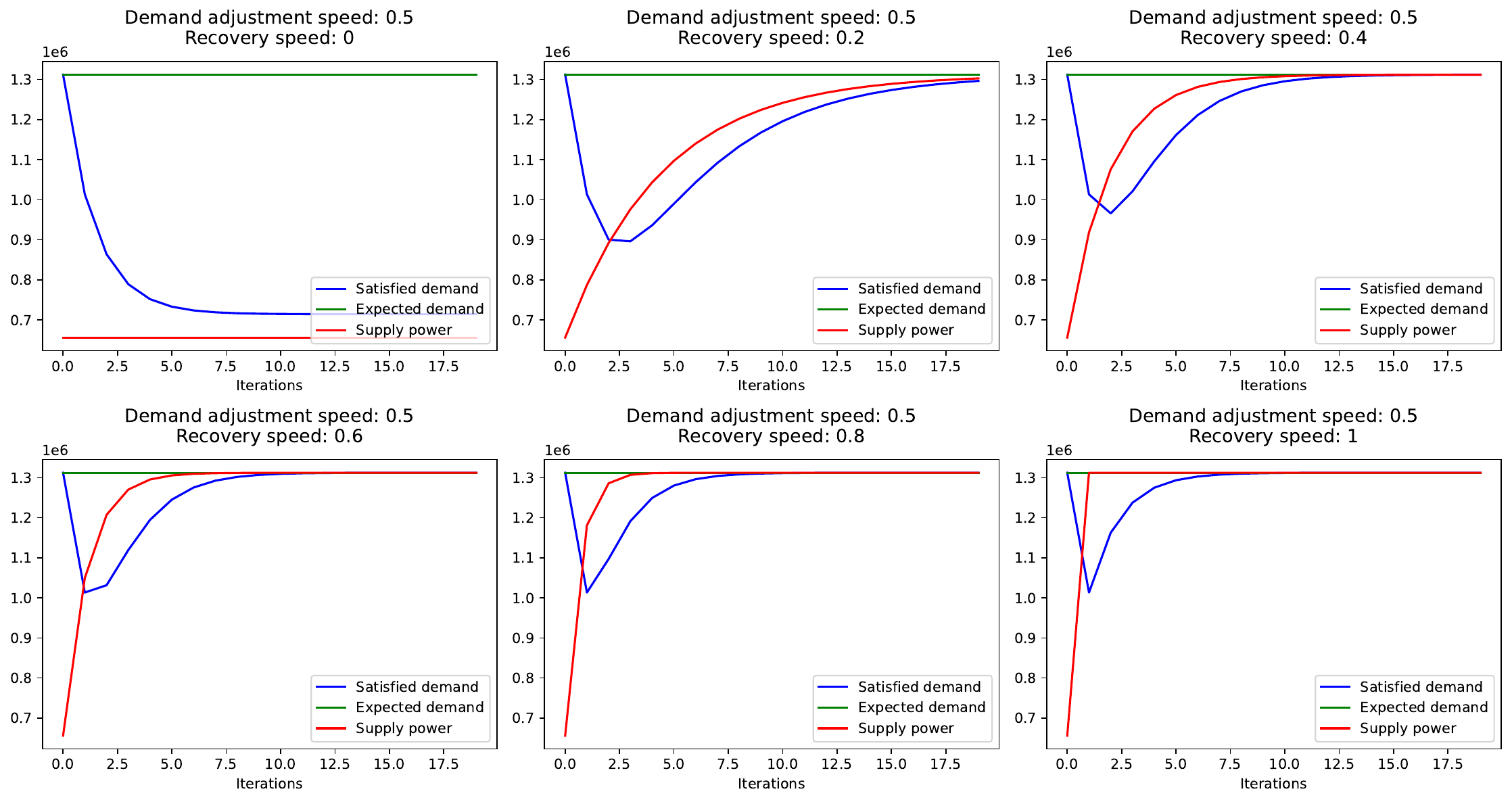}
        \caption{\textit{The impact of supply recovery speed with a fixed demand adjustment speed of 0.5}}\label{fig:recovery_2_sub}
    \end{subfigure}
    \caption{\textit{Recovery dynamics under different scenarios}}
\end{figure}

In \cref{fig:recovery_1_sub} and \cref{fig:recovery_2_sub}, we simulate a 50\% shock to the mining industry and track three variables: satisfied demand (demand not immediately adjusted), expected demand (assumed constant at the original level, reflecting long-term targets), and supply power (which gradually recovers). The shock reduces supply power to 50\%, after which it steadily returns to the pre-shock level. Satisfied demand gradually converges to supply power, producing a hump-shaped pattern that captures the dynamics of shock propagation and economic recovery.

In \cref{fig:recovery_1_sub}, we hold the supply recovery speed constant while varying the demand adjustment speed to illustrate how propagation unfolds. In contrast, \cref{fig:recovery_2_sub} keeps the demand adjustment speed fixed and varies the recovery speed to demonstrate differences in recovery dynamics.

When supply recovery speed is fixed at 10\%, increasing the demand adjustment speed makes the economy more vulnerable to the shock. When adjustment speed is zero, demand is not sensitive to supply, thus supply shock will not affect the demand level. A high adjustment speed suggests limited inventory or a lack of flexibility in sustaining demand, leading to a sharp drop in satisfied demand and more severe economic disruption. Conversely, with slower adjustment, demand declines more gradually, allowing supply recovery to mitigate the shock before inventories are exhausted. As a result, the overall economic impact is less severe, and recovery begins earlier.

In \cref{fig:recovery_2_sub}, we fix the demand adjustment speed at 0.5 and vary the supply recovery speed. As the recovery speed increases, both the severity and duration of the economic impact decrease. When supply recovery speed is zero, demand will gradually converges to the supply level without recovery. A faster supply recovery speed steepens the supply power curve, allowing demand to converge more quickly toward expected demand. This results in a shorter and less pronounced disruption to the economy.
A word on the units of these speeds is in order. The units are once again relative. Over one iteration, a recovery speed of 0.1 implies that the supply recovers to 10\% of its original value. 

\section{Conclusion}

This paper presents a framework for analyzing how supply shocks propagate through input-output (IO) networks. The Priority with Constraint algorithm provides a minimum supply guarantee, creating a buffering effect and highlights critical shock propagation pathways that contribute to systemic risk. Additionally the dynamic modeling approach demonstrates the critical role that demand adjustment speed and supply recovery rates play in determining the severity and duration of economic shocks. Specifically, rapid demand adjustments amplify economic vulnerability, whereas accelerated supply recovery significantly reduces disruption.

While IO modeling provides substantial insight into economic dynamics, the aggregated and coarse-grained nature of the underlying data presents limitations. Therefore, connecting this aggregated perspective with more detailed, fine-grained data, ideally at the single firm level, is essential. Furthermore, the time scales of adjustments and impacts at the sectoral level - captured by IO models - often differ significantly from those experienced at the individual firm level. 

Despite these limitations, IO modeling remains a valuable starting point for understanding crucial economic dynamics. Looking ahead, this framework offers valuable applications for financial risk modeling, particularly in addressing physical risks arising from events such as climate change, geopolitical conflicts, demographic shifts, or technological disruptions (e.g., AI). Within the context of financial risk assessment, the framework can help measure portfolio risk by identifying and assessing exposure to companies vulnerable to physical events; evaluate return correlation by understanding new commonalities in asset returns driven by unprecedented physical risks; recast sector and industry factors by adapting factor models to reflect exposures and correlations during physical disruptions not captured by historical data; develop product-level factors based on products, fundamental building blocks for both physical risk assessment and industry classification. Finally, it also opens the door for the creation of physical risk scenarios for financial stress testing by explicitly modeling physical disruptions.
\bibliography{references}

\begin{thebibliography}{6}%
\makeatletter
\providecommand \@ifxundefined [1]{%
 \@ifx{#1\undefined}
}%
\providecommand \@ifnum [1]{%
 \ifnum #1\expandafter \@firstoftwo
 \else \expandafter \@secondoftwo
 \fi
}%
\providecommand \@ifx [1]{%
 \ifx #1\expandafter \@firstoftwo
 \else \expandafter \@secondoftwo
 \fi
}%
\providecommand \natexlab [1]{#1}%
\providecommand \enquote  [1]{``#1''}%
\providecommand \bibnamefont  [1]{#1}%
\providecommand \bibfnamefont [1]{#1}%
\providecommand \citenamefont [1]{#1}%
\providecommand \href@noop [0]{\@secondoftwo}%
\providecommand \href [0]{\begingroup \@sanitize@url \@href}%
\providecommand \@href[1]{\@@startlink{#1}\@@href}%
\providecommand \@@href[1]{\endgroup#1\@@endlink}%
\providecommand \@sanitize@url [0]{\catcode `\\12\catcode `\$12\catcode `\&12\catcode `\#12\catcode `\^12\catcode `\_12\catcode `\%12\relax}%
\providecommand \@@startlink[1]{}%
\providecommand \@@endlink[0]{}%
\providecommand \url  [0]{\begingroup\@sanitize@url \@url }%
\providecommand \@url [1]{\endgroup\@href {#1}{\urlprefix }}%
\providecommand \urlprefix  [0]{URL }%
\providecommand \Eprint [0]{\href }%
\providecommand \doibase [0]{https://doi.org/}%
\providecommand \selectlanguage [0]{\@gobble}%
\providecommand \bibinfo  [0]{\@secondoftwo}%
\providecommand \bibfield  [0]{\@secondoftwo}%
\providecommand \translation [1]{[#1]}%
\providecommand \BibitemOpen [0]{}%
\providecommand \bibitemStop [0]{}%
\providecommand \bibitemNoStop [0]{.\EOS\space}%
\providecommand \EOS [0]{\spacefactor3000\relax}%
\providecommand \BibitemShut  [1]{\csname bibitem#1\endcsname}%
\let\auto@bib@innerbib\@empty
\bibitem [{\citenamefont {Leontief}(1936)}]{leontief1936quantitative}%
  \BibitemOpen
  \bibfield  {author} {\bibinfo {author} {\bibfnamefont {W.~W.}\ \bibnamefont {Leontief}},\ }\bibfield  {title} {\bibinfo {title} {Quantitative input and output relations in the economic systems of the united states},\ }\href@noop {} {\bibfield  {journal} {\bibinfo  {journal} {The review of economic statistics}\ ,\ \bibinfo {pages} {105}} (\bibinfo {year} {1936})}\BibitemShut {NoStop}%
\bibitem [{\citenamefont {Leontief}(1986)}]{leontief1986input}%
  \BibitemOpen
  \bibfield  {author} {\bibinfo {author} {\bibfnamefont {W.}~\bibnamefont {Leontief}},\ }\href@noop {} {\emph {\bibinfo {title} {Input-output economics}}}\ (\bibinfo  {publisher} {Oxford University Press},\ \bibinfo {year} {1986})\BibitemShut {NoStop}%
\bibitem [{\citenamefont {Weber}\ \emph {et~al.}(2024)\citenamefont {Weber}, \citenamefont {Lara~Jauregui}, \citenamefont {Teixeira},\ and\ \citenamefont {Nassif~Pires}}]{weber2024inflation}%
  \BibitemOpen
  \bibfield  {author} {\bibinfo {author} {\bibfnamefont {I.~M.}\ \bibnamefont {Weber}}, \bibinfo {author} {\bibfnamefont {J.}~\bibnamefont {Lara~Jauregui}}, \bibinfo {author} {\bibfnamefont {L.}~\bibnamefont {Teixeira}},\ and\ \bibinfo {author} {\bibfnamefont {L.}~\bibnamefont {Nassif~Pires}},\ }\bibfield  {title} {\bibinfo {title} {Inflation in times of overlapping emergencies: Systemically significant prices from an input--output perspective},\ }\href@noop {} {\bibfield  {journal} {\bibinfo  {journal} {Industrial and Corporate Change}\ }\textbf {\bibinfo {volume} {33}},\ \bibinfo {pages} {297} (\bibinfo {year} {2024})}\BibitemShut {NoStop}%
\bibitem [{\citenamefont {Pichler}\ and\ \citenamefont {Farmer}(2022)}]{pichler2022simultaneous}%
  \BibitemOpen
  \bibfield  {author} {\bibinfo {author} {\bibfnamefont {A.}~\bibnamefont {Pichler}}\ and\ \bibinfo {author} {\bibfnamefont {J.~D.}\ \bibnamefont {Farmer}},\ }\bibfield  {title} {\bibinfo {title} {Simultaneous supply and demand constraints in input--output networks: the case of {C}ovid-19 in {G}ermany, {I}taly, and {S}pain},\ }\href@noop {} {\bibfield  {journal} {\bibinfo  {journal} {Economic Systems Research}\ }\textbf {\bibinfo {volume} {34}},\ \bibinfo {pages} {273} (\bibinfo {year} {2022})}\BibitemShut {NoStop}%
\bibitem [{\citenamefont {Bartolucci}\ \emph {et~al.}(2025)\citenamefont {Bartolucci}, \citenamefont {Caccioli}, \citenamefont {Caravelli},\ and\ \citenamefont {Vivo}}]{bartolucci2025upstreamness}%
  \BibitemOpen
  \bibfield  {author} {\bibinfo {author} {\bibfnamefont {S.}~\bibnamefont {Bartolucci}}, \bibinfo {author} {\bibfnamefont {F.}~\bibnamefont {Caccioli}}, \bibinfo {author} {\bibfnamefont {F.}~\bibnamefont {Caravelli}},\ and\ \bibinfo {author} {\bibfnamefont {P.}~\bibnamefont {Vivo}},\ }\bibfield  {title} {\bibinfo {title} {Upstreamness and downstreamness in input--output analysis from local and aggregate information},\ }\href@noop {} {\bibfield  {journal} {\bibinfo  {journal} {Scientific Reports}\ }\textbf {\bibinfo {volume} {15}},\ \bibinfo {pages} {2727} (\bibinfo {year} {2025})}\BibitemShut {NoStop}%
\bibitem [{\citenamefont {Sharma}\ \emph {et~al.}(2020)\citenamefont {Sharma}, \citenamefont {Bouchaud}, \citenamefont {Gualdi}, \citenamefont {Tarzia},\ and\ \citenamefont {Zamponi}}]{sharma2020v}%
  \BibitemOpen
  \bibfield  {author} {\bibinfo {author} {\bibfnamefont {D.}~\bibnamefont {Sharma}}, \bibinfo {author} {\bibfnamefont {J.-P.}\ \bibnamefont {Bouchaud}}, \bibinfo {author} {\bibfnamefont {S.}~\bibnamefont {Gualdi}}, \bibinfo {author} {\bibfnamefont {M.}~\bibnamefont {Tarzia}},\ and\ \bibinfo {author} {\bibfnamefont {F.}~\bibnamefont {Zamponi}},\ }\bibfield  {title} {\bibinfo {title} {V-, {U}-, {L}-, or {W}-shaped recovery after {COVID}? {I}nsights from an {A}gent {B}ased {M}odel},\ }\href@noop {} {\bibfield  {journal} {\bibinfo  {journal} {PLoS One}\ }\textbf {\bibinfo {volume} {16}} (\bibinfo {year} {2020})}\BibitemShut {NoStop}%
\end{thebibliography}%
\appendix

\section{Rationing Algorithms}\label{app:ration_algorithm}

We discuss the details of the various rationing algorithms used in the analysis here. 

\subsection*{Proportional Rationing}

In the proportional rationing algorithm, we have:
\begin{equation}
r_i[t] = \frac{x_i^{\text{max}}}{d_i[t]}
\end{equation}

where:

\begin{itemize}
    \item \(r_i[t]\) is the rationing factor for industry \(i\) at time step \(t\),
    \item \(x_i^{\text{max}}\) is the maximum output that industry \(i\) can produce (based on supply constraints),
    \item \(d_i[t]\) is the total demand for goods from industry \(i\) at time \(t\).
\end{itemize}

In this proportional algorithm, the rationing factor \(r_i[t]\) tells us the proportion of total demand \(d_i[t]\) that can be satisfied given the supply constraint \(x_i^{\text{max}}\). If \(r_i[t] < 1\), the industry can only meet part of its demand, and the resources are distributed proportionally among all customers. This approach assumes no distinction between intermediate demand (other industries) and final demand (consumers), treating all demand equally.

\subsection*{Mixed Proportional/Priority Rationing}

In the mixed proportional/priority algorithm, we have: 

\begin{equation}
r_i[t] = \frac{x_i^{\text{max}}}{\sum_j a_{ij} d_j[t]}
\end{equation}
where $\sum_j a_{ij} d_j[t]$ represents the total intermediate demand for industry \(i\)'s products, meaning the sum of all demand from other industries (\(j\)) that use \(i\)'s outputs as inputs.

This equation only takes intermediate demand into account when determining the rationing factor. In other words, the algorithm gives priority to intermediate demand—i.e., industries that need inputs from industry \(i\) to produce their own goods. Final demand from consumers (households, government, exports) is deprioritized. The rationing factor \(r_i[t]\) here is based solely on the needs of other industries, meaning that final consumers will only receive goods after the intermediate demand has been met.

\subsection*{Priority Rationing (Largest First)}

In the priority rationing algorithm, we have: 

\begin{equation}
r_{ij}[t] = \frac{x_i^{\text{max}}}{\sum_{n \in h_{ij}} a_{in(j)} d_{n(j)}[t]}
\end{equation}

where $h_{ij}$ represents the ranked ordering of customers based on the size of their demand, and $\sum_{n \in h_{ij}} a_{in(j)} d_{n(j)}[t]$ is the sum of demands from the largest customers to the smallest, reflecting a priority ordering based on demand size.

This equation differs from the others by incorporating priority based on customer size. Industries rank their customers and serve the largest first. The rationing factor $r_{ij}[t]$ is calculated by determining how much of the total demand from prioritized customers can be met, starting with the largest. Once the largest customer's demand is met (if possible), resources are allocated to the next, and so on. Smaller customers may receive little or no supply if the larger ones consume most of the available output.

\subsection*{Priority with Constraint Rationing}
The largest first algorithm prioritizes large customers during supply constraints, while the modified largest first algorithm applies a minimum supply buffer $\mathcal{S}_{min}$, reducing impact on minor customers. $\mathcal{S}_{min}$ is the least amount of input that a supplier is willing or able to provide to its downstream industries. The modified approach ensures that small customers are less severely affected.

\begin{align}
    r_{ij}[t] &= \frac{\max(x_i^{\max} - \sum_{n \in h_{ij}} a_{in(j)} d_{n(j)}[t], \ \mathcal{S}_{min})}{d_{n(j)}[t]}
\end{align}

\section{Input-Output Algorithm}\label{app:io_algorithm}

All of the algorithms presented previously are based on the calculation of bottleneck, which is the minimum supply power of all suppliers given an industry. Since an industry's supply power is constrained by all of its suppliers, it makes sense to have the minimum supply power of all suppliers as its constraint. Different algorithms only differ in how they calculate bottlenecks.

To illustrate, we first assume the bottleneck is calculated and demonstrate the propagation of shock (common to all algorithms). Then, we discuss how the bottleneck is calculated.

\begin{itemize} 
    \item \textbf{The Common Influence Given Bottlenecks}
The process is an adjustment of demand given a supply shock. Firstly, the industry experiences the bottleneck imposed by all its suppliers. Then, societal demand decreases due to insufficient supply. Through this iteration, supply and demand will match again.

\item \textbf{Industry Suffering from Bottleneck}
Given each industry $i$'s supply constraint $r_{ji}$ (where $j$ are suppliers of industry $i$), the bottleneck $s_i$ is the minimum constraint of all suppliers: $s_{i} = \min(r_{ji}, 1)$.

\item \textbf{Bottleneck Constraining Supply Power}
The bottleneck $s_i$ constrains the supply of industry $i$ to downstream industries, scaling down the supply $x$.

\item \textbf{Final Consumer Demand Shrinks for Market Clearance}
For market equilibrium, the final consumer demand for industry $i$, $f_i$, equals the supply $x_i$ minus intermediary demands.

\item \textbf{Society's Demand Adjusts Accordingly}
According to Leontief's input-output model, a decline in consumer demand will propagate to industry demand in the next period, $d_i[t+1]$.
\end{itemize}

\subsection*{Bottleneck Calculation Methodologies}

The difference in algorithms arises in calculating bottlenecks under supply shock. For an industry $i$, the bottleneck is defined by supply constraints from all its suppliers. 

We can summarize the different algorithms through the following steps:

\begin{align}
r_i[t] &= f(x_i^{\max}, ed_i[t]) \\
s_i[t] &= \min \left\{ r_{ij}[t], 1 \right\}, \\
x_i[t] &= \min \left\{ x_i^{\max}, s_i[t] d_i[t] \right\}, \\
f_i[t] &= \max \left\{ x_i[t] - \sum_{j} a_{ij} x_j[t], 0 \right\}, \\
d_i[t + 1] &= \sum_{j} l_{ij} f_j[t].
\end{align}

This process first calculates the bottleneck $s_{i}$ of each industry $i$, which is represented by different rationing algorithms via the function $f$ here. Then, the supply power is affected $x_{i}$, which decreases demand due to market clearing conditions. 

\section{Recovery Mechanism}\label{app:recovery_mechanism}

We make two modifications to the process outlined in \cref{app:io_algorithm}. First, demand will not adjust to supply constraints quickly, but will gradually decrease with an adjustment speed (\textit{Sticky demand part}). Second, supply will recover with a recovery factor (\textit{Recovery of actual supply}). The recovery motivation of demand is essentially a backup force for the recovery of supply, which is not decisive (simply put, the economy has a motivation to return to its original state, so as supply recovers, demand also responds).

Secondly, we use expected demand $d_{i}^{exp}$ to calculate the bottleneck instead of actual demand, which is always lower than expected demand but has a tendency to recover to targeted demand (the original level).

The modified process is:

\begin{align}
\text{Recovery of expected demand:} \quad & d_i^{exp}[t] = (1 - \alpha) d_i[t] + \alpha d_{i}^{exp}, \\
\text{Rationing algorithm:} \quad &
\begin{aligned}
    r_i[t] &= f(x_i^{\max}, d_{i}^{exp}[t]), \\
    s_i[t] &= \min \{r_{ij}[t], 1\}, \\
    x_i[t] &= \min \{x_i^{\max}, s_i[t] d_i[t]\}, \\
    f_i[t] &= \max \{x_i[t] - \sum_j a_{ij} x_j[t], 0\},
\end{aligned} \\
\text{Sticky demand adjustment:} \quad & d_i[t+1] = (1 - \beta) d_i[t] + \beta \sum_j l_{ij} f_j[t], \\
\text{Supply capacity recovery:} \quad & x_i^{\max}[t+1] = (1 - \gamma) x_i^{\max}[t] + \gamma x_i^{\text{ori}}.
\end{align}

\end{document}